# A Private and Unlinkable Message Exchange Using a Public bulletin board in Opportunistic Networks

Ardalan Farkhondeh

Department of Information and Communications Engineering

Autonomous University of Barcelona

ardalan.farkhondeh@e-campus.uab.cat

**ABSTRACT**

We plan to simulate a private and unlinkable exchange of messages by using a Public bulletin board and Mix networks in Opportunistic networks. This Opportunistic network uses a secure and privacy-friendly asynchronous unidirectional message transmission protocol. By using this protocol, we create a Public bulletin board in a network that makes individuals send or receive events unlinkable to one another [18]. With the design of a Public bulletin board in an Opportunistic network, the clients can use the benefits of this Public bulletin board in a safe environment. When this Opportunistic network uses the protocol, it can guarantee an unlinkable communication based on the Mix networks. The protocol can work with the Public bulletin board exclusively with acceptable performance. Also, this simulation can be used for hiding metadata in the bidirectional message exchange in some messengers such as WhatsApp. As we know, one of the main goals of a messenger like WhatsApp is to protect the social graph. By using this protocol, a messenger can protect social graph and a central Public bulletin board.

## 1. INTRODUCTION

### 1.1 General Objective

We plan to simulate a private and unlinkable exchange of messages by using a Public bulletin board and Mix networks in Opportunistic networks. To do this, we have designed an Opportunistic network that uses a secure and private integrated protocol. We can use this project in mail providers or messengers. In our proposal, clients can send their messages in a safe and secure environment. Users can use the advantage of a public bulletins board and save their messages in a database securely. Two nodes allow for communication in an Opportunistic network so that data transference takes place by using Mix networks and Public bulletin boards in a way that the contact address is not revealed to the participants in the communication [18].

By using Mixer nodes in our proposal, we can make a non-tracing Opportunistic network, and a client can send and receive a message anonymously. In this Opportunistic network, malicious users cannot rebuild social structure networks.

designed indirect and non-tracing communication. In a nutshell, we can say that our primary goal of simulation of this proposal is creating Opportunistic networks by using Public bulletin boards and Mix networks in a non-tracing space.

### 1.2 Document Structure

This document consists of the Abstract and Introduction sections and the Related works section. The Related works section includes three main subsections: Definition of Opportunistic networks, Anonymity in Opportunistic networks and Mix networks.

In the first subsection, the document offers a definition for Opportunistic networks and discusses five routing methods in Opportunistic networks as well as the Architecture of Opportunistic networks. This section tries to familiarize the reader with the basic concepts of Opportunistic networks.

In Anonymity in Opportunistic networks section have two subsections. In the first part, the documents introduced a definition of anonymity in the opportunistic network. Also, this part tries to familiarization the readers with some basic concept of Anonymity in Opportunistic networks. In the second subsection the document introducing some anonymous routing protocols and explain all of them briefly.

The third subsection explains about Mix networks. Mix networks subsection has three parts that including a definition of Mix networks, the description of mixer nodes and privacy issues in Mix networks. Since our simulation model has used Mix networks, this document needs to explain the structure and general concepts of Mix networks. In this subsection, we review Mix networks and explain about the node's behaviour in them. The section illustrates how Mix networks can create paths by using mixer nodes. This subsection also discusses privacy issues in Mix network. Furthermore, we offer some suggestion for improving privacy in Mix networks in the end.

The fourth section is the Proposal section. It includes two main subsections. The first one discusses the structure of Public bulletin boards and their advantages. It offers a detailed explanation of different parts of the Public bulletin board structure. We clarify that a Public bulletin board can have that include Write method, Read method and response method. The second subsection of the proposal discusses client protocol for the transmission of packets between nodes and a Public bulletin board by Mix networks.

The fifth section of the document is the Experimentation section. It includes four subsections. The first subsection describes the simulation environment. In general, in this subsection, we talk about the ONE simulator and describe the differences between java functions in the ONE simulator. The next subsection is Scenario in which we explain about two scenarios run in our simulation. Also, we discuss the differences and similarities of those scenarios. The next subsection is the ONE modification in which we explain all the changes that we made on some functions of the ONE simulator. At the end of Experimentation section, we review the results of both scenarios and discuss all the metrics we have used. Next, we compare both scenarios and present a conclusion on the results.

The final two sections of the document are the conclusion of the whole document and the bibliography.

## 2. RELATED WORK

### 2.1 Opportunistic networks

#### 2.1.1 Definition and characteristics

One subclass of the Delay-Tolerant Networks is Opportunistic networks. An Opportunistic network has intermittent communication opportunities in a way that there ever exists an end-to-end path between the sources and the destinations. In an Opportunistic network, link performance is too variable. We cannot use TCP/IP protocol in an Opportunistic network, because of an end-to-end path created between the source and the destination for a short period of time and its unpredictability. An Opportunistic network can use node mobility to exploits and local forwarding for data transference, which can be an excellent suggestion to forfeit the lack of TCP/IP protocols [1]. Each node can store and carry data until these nodes have opportunistic contact. The nodes can forward the data when they have opportunistic contact with other nodes. In general, entire messages transferred from a node's buffer to another node's buffer until the messages reach their destination by using a path.

In general, we can mention three main characteristics of Opportunistic networks. These characteristics include flexibility for different environments, low budget and easy implementation. These characteristics help us develop Opportunistic networks quickly. In what follows, we explain each characteristic in more details.

The first characteristic of Opportunistic networks is their flexibility. Opportunistic networks can include a large number of nodes that are always on the move. The movement of nodes can be used to convey messages.

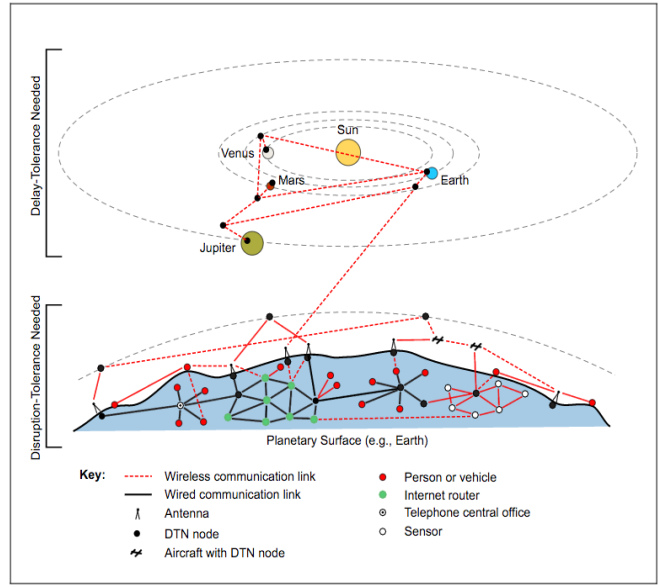

**Figure 1. An example of an Opportunistic network. Adapted from [4]**

The nodes can share their messages with their neighbour nodes. Since nodes are moving, they can transfer the messages from one node to other groups of nodes until the messages arrive at the destination node. The movement of nodes eliminates the need to install the specific infrastructure for transmission of messages in Opportunistic networks. Network designer can design different protocols on movement devices easily, and these protocols can smoothly adapt to the needs of different environments in order to convey messages. Advantages of using these networks include their utilizing a variety of protocols, high versatility with different environments and their use of the movement of nodes for transmission which Opportunistic networks highly flexible.

Therefore, we do not need to spend much money to install specific infrastructures. For this reason, we can implement Opportunistic networks without any government funding most of the time. To do so, we only need to use some cheap or free applications on local devices which are capable of sending and receiving messages in one area. Due to the affordability of Opportunistic networks, using them in poor or remote areas has been welcomed.

The last characteristic of Opportunistic networks is related to their easy implementation. Network designer needs to define some protocols and some programming in the form of an application, after that, this application will be installed on the local devices in one area. Thus, this process can be done in a short time. We can imagine an area which has lost all of the network infrastructures by flooding or war and this area needs implementation of a network in a short period of time. Obviously, in such situations, one of the most reasonable choices are Opportunistic networks because of their quick implementation.

**2.1.2 Two Examples of Opportunistic network**

In Figure 1, a simple example of the operation of an Opportunistic network is displayed. In this figure, we can see different nodes of people and vehicles. These nodes include peoples, cars and satellites that form networks based on the different infrastructures on earth and sky. Some nodes use wired communication to form the networks; for instance, these nodes connect to a central telephone office with wired connections. Also, we can see that some nodes create networks by using a wireless connection. We also see that some sensors or some nodes have established networks based on internet routing. This figure well shows a tolerant network based on a set of heterogeneous networks. In this figure, all subset networks eventually connect into DTN nodes. A DTN node is a node which has the potential to tolerate the delay in message transmission paths.

As shown in Figure 2, this Opportunistic network consists of three groups (A, B, and C) of nodes located in separate locations. Each group consists of two types of node. The first type of node merely consists of group members and plays a role in transmitting messages inside their location. These normal nodes cannot transfer the messages to the outside, but these nodes can transfer ordinary messages between members. Other types of nodes are intermediate nodes which can send obfuscated messages between different groups of nodes in different locations [3].

As shown in this figure, the node U1 generates the message and sends it to the node U2, which is an intermediate node. Node U2 transfers the message from the group location A to the group location B. Node U2 moves to the group location B, and then node U2 publishes the messages among the second-nodal group members. After being published, the message arrives at the node U4, which is an intermediate node. Node U4 moves to the next location and publishes the message to the group C. After the message reaches U5, we have established a perfect connection between the nodes in different places. Now, the node U1 can send the message through the intermediate nodes to the node U5. The node U5 can send the message through antenna and route through the various routers to a service provider centre. Thus, this figure shows an implementation that a node that can by using intermediate nodes in different locations send a message to a long-distance service provider.

Figure 2 shows that we can make a path between two long-distance nodes. Moreover, due to the heterogeneity of the path, the Opportunistic network may be associated with many disturbances. Some disturbances may cause the Opportunistic network to fail the message along the transferring path.

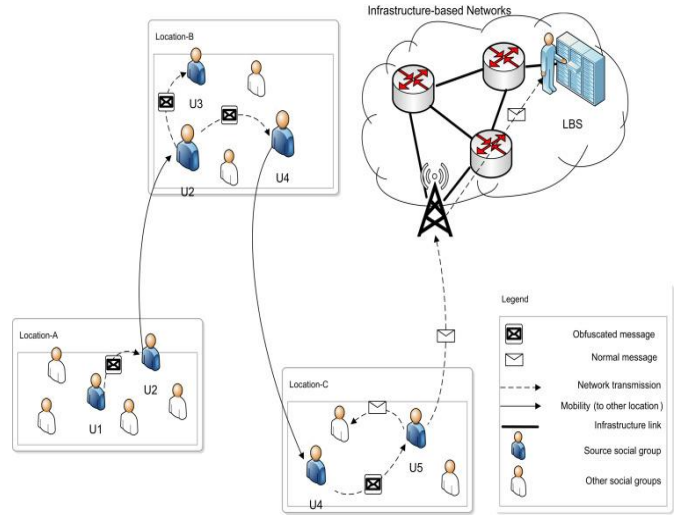

**Figure 2. An example of an Opportunistic network in real life. Adapted from [5]**

Network designers need to plan nodes to solve such kinds of problems in different situations. It illustrates a condition in which the node may not have enough space to buffer a message. This node may not have enough space to buffer a message. Therefore, we propose that buffers be shared between neighbouring nodes. In this case, if a node encounters a space shortage, it can use the neighbouring node buffers. This suggestion can be an example of network flexibility in node planning to solve problems.

**2.1.3 Classification of routing in Opportunistic networks**

We can classify routing algorithms based on the data forwarding behaviour in the Opportunistic networks in five main categories. These categories include Direct Transmission, Flooding Based, Prediction Based, Coding Based, and Context-based routing [2].

*A. Direct Transmission*

In this algorithm of routing, each node generates a packet for sending to the destination. To do so, they store the in their buffer at first. The packets remain in the buffers until the nodes meet up the node destination.

The nodes can forward the message directly to their desired destination, and they just forward a single copy of the message. Therefore, nodes must wait for an opportunity to meet their desired destination and then send their message directly. The disadvantage of this method is that it is unclear when a sender and receiver can meet each other; thus, this routing algorithm has an unbounded delivery delay. This algorithm has an advantage since it needs one transmission to forward a single copy of the packets.

*B. Flooding Based routing*

In Flooding Based algorithm of routing, the nodes generate a multi-copy of packets and inject them to the Opportunistic network. This process continues until the packets arrive at the destination — flooding at source nodes classified into two categories, which includes controlled and uncontrolled categories. In the controlled category, replication of the packets is unlimited to reduce the long delivery delay. In the controlled category, the opportunistic networks set a limit for replication of the packets to reduce the network contention.

*C. Prediction Based routing*

It should be noted that the Flooding Based routing sometimes receives the amount of traffic overhead, and Prediction Based algorithm introduces a way to reduce the traffic. In Prediction Based algorithm, the Opportunistic network predicts and calculates the nodes' behaviour according to their contact history, and the Opportunistic network can decide to predict the best probabilistic routing. If a neighbour node has higher delivery predictability value, a sender node forwards its packet to this node.

*D. Coding Based routing*

In Coding Based algorithm of routing, the Opportunistic network changes the format of all the packets before transmission. Coding based routing embeds additional information in blocks of the packets. If the receivers want to reconstruct the original packets, they must open a certain number of blocks successfully.

*E. Context based routing*

The Context-Based algorithm of routing tries to improve the Prediction based algorithm by utilizing the context information. Sometimes, Prediction Based algorithm fails to predict the right paths; thus, the Opportunistic network uses the context information that can refine the wrong prediction and improve the delivery ratio.

**2.1.4 Architecture of opportunistic networks**

It is important to know than an Opportunistic network typically includes separate partitions. These partitions are called regions. By using devices in different regions, Opportunistic networks try to store-carry-forward the packet and make connections between heterogeneous regions. By using a bundle layer, the intermediate nodes implement the store-carry forward message switching mechanism. The act of nodes is defined by the bundle layer. It should be noted that that nodes can perform three acts, as a host, a router and a gateway. The bundle layer of the router can store, carry and forward the packets between nodes in the same region. The bundle layer of the gateway can store, carry and forward the packets between different regions. The gateway can also be a host, optionally [3].

**2.2 Anonymity in Opportunistic networks**

**2.2.1 Introduction to Anonymous communication**

The purpose of anonymous communication is to communicate between senders and receivers who know each other, while observers and network entities do not recognize the identities of the senders and receivers [6]. Malicious users cannot trace the traffics, nor can they discover the identities of the sender and receiver.

As this document discussed, the Opportunistic networks can be used in mobile communications, network communications in remote areas, intelligent transportation and many other types. However, in each different implementation of Opportunistic networks, we must always protect the privacy. The information potential of each node can be a threat and cause serious security problems. We should find a way to protect the information and prevent from tracing it by dangerous people who can cause irreparable problems. One of the best ways to protect the nodes' privacy is to keep their anonymity. Therefore, discussing the anonymity of nodes and protecting personal information is always a severe challenge in Opportunistic networks.

Opportunistic networks must strive to preserve critical information, including names, messages, places and other types of information. The formation of social graphs and social relationships that are created by the nodes meeting must be preserved. The anonymity of the spheres on the Opportunistic network guarantees that it can prevent from the restructuring of social graphs from malicious users [7].

Keeping the node's anonymity by the Opportunistic networks does not mean that any of the node's information cannot be shared because in many cases, they should also share some of the node's information with other nodes. Then Opportunistic networks should be able to balance between the disclosure of the node's identity and the necessary information. The Opportunistic network must try to prevent malicious users that try to use confidential information. For this purpose, an Opportunistic network needs an integration mechanism to protect privacy. An Opportunistic network needs an integrated security mechanism across the entire network environment, and the network must increase the level of trust between nodes. With an integrated security system, the network can better preserve the social graph.

An Opportunistic network can do two general possibilities for authentication. Before the Opportunistic network publishes all identifying information of nodes, there is no social graph. The first possibility is to allow the nodes to place all of their identity information on the Opportunistic network. In this case, the Opportunistic network is very vulnerable. Therefore it must consider a policy among the nodes that the network can protect the form of the social graph and keep important information at the same time. If an Opportunistic network publishes all the information about all nodes, creating a social graph is meaningless. Another possibility is that the network hides all the nodes' information. Moreover, this policy cannot be realistic. It does not allow for the formation of social graphs, and this network cannot be very efficient [9].

This paragraph suggests implementing an integrated security mechanism. One of the solutions for authentication with anonymity is the use of digital signatures on packets. The use of a unique signature in the router enables the Opportunistic network to identify identities from packets and send messages in a more efficient way. Also, the network can better use resources. Also, after the security authorization was granted by routers, the routers add individual licenses to the packets. All this requires additional computing on the Opportunistic network; instead, it makes for integrated security on the network [8,11].

**2.2.2 Introducing some anonymous routing protocols**

*A. Onion routing*

One famous DTN routing protocol for anonymity is onion routing. Onion routing with layered encryption on packets by using different secret keys preserves anonymity. The peeling off of each layer in DTNs is done by the corresponding secret key. This protocol remains the anonymous connection between the sender, receivers and the final destination, so the connection between them becomes untraceable [10]. The paths in onion routing are made by some onion routers. Each onion router only recognizes the previous and the next router, and none of them knows the first sender and the final destination.

*B. ALAR*

ALAR is an anonymity protocol for improving level privacy by combining the ideas of on-demand routing, identity-free routing, and neighbourhood traffic. The main idea of ALAR is to fragment a message to K segment and to send each segment to at least N neighbours [12]. In ALAR protocol, the malicious user receives many copies of a packet from different routers at a different time so that if they want to trace the packets of the traffic probably cannot find the sender and receivers.

*C. MASK*

MASK is an anonymous on-demand protocol. Neighbouring nodes share a pairwise secret key between each other to establish a neighbourhood authentication without revealing their identities. With node's secret key, MASK can forward and route packets without disclosing of identities of nodes [12]. However, we should know that malicious users can localize the sender's position, which is a weakness.

*D. ASR*

ASR is a secure routing protocol that only encrypts a small piece of a packet instead of encrypting the whole packet. ASR only encrypts a small part of a packet which contains the information on the identity of the sender and the receiver. It should also be noted that each relay node only verifies that small encrypted piece of a packet. The forwarding process functions by sharing a key between any two consecutive nodes until the message arrives at the destination [12].

*E. Privhab*

Privhab uses homomorphic encryption with a protocol that functions according to the location of the nodes (Homomorphic encryption can encrypt data with additional evaluation capability, and Homomorphic encryption can use that additional evaluation value for computing over encrypted data without using the secret key). This method seeks nodes that are within the boundary range and use them to select the probability of intermediate nodes. Also, this method uses long-term predictions to select the intermediate nodes, which increases network efficiency. Privhab uses an innovative way to protect privacy. Instead of using the actual user name, this method uses aliases to keep the identities of the nodes intact [13].

*F. Eprivo*

Eprivo uses Homomorphic encryption for keeping privacy and anonymization. The Homomorphic encryption had used for avoiding reveals information corresponding to the node's neighbouring graph in Eprivo. When two nodes meet each other, they do not share their routing metrics. They only compare routing metrics by using Homomorphic encryption. The Eprivo have defined two main methods to keep anonymous of the nodes. These two methods include Neighborhood-randomization and Binary-anonymization [14].

**2.3 Mix networks**

**2.3.1 Defination of Mix networks**

Mix networks are introduced as anonymous communication protocols that can prevent the tracing of senders and recipients. A Mix network uses the mixer nodes to prevent malicious users from tracking messages. As we know, Opportunistic networks can build paths to transmit messages by using intermediate nodes. Mix networks can use mixer nodes between intermediate nodes to construct the paths.

Moreover, Mix networks have a high degree of flexibility, the same as Opportunistic networks. For this reason, we can consider Mix networks for different locations and environments. However, the term "Mix networks" is often referred to as non-traceable secure networks. Mix networks are very suitable for particular environments where information preservation is of great importance. For example, a government can use a Mix network to hold an important election in a country, where the disclosure of information or information tracking can lead to many problems.

**2.3.2 Defination of Mixer nodes**

Mixer nodes store all the received messages sent by the surrounding nodes in their buffers. As we see in figure 3, when the mixer node receives messages, they do not send messages immediately to other nodes. The mixer nodes change the order of all the messages before sending them to the next destination. These changes randomly occur in the mixer node buffer. We should know that the volume of Mix network traffic directly depends on the number of mixer nodes. The higher the number of the mixer nodes grows, the higher the amount of traffic becomes equally [15]. After the messages shuffled in a mixed column in a mixer node, they are sent to the nodes around the mixer nodes.

All messages in Mix networks are encrypted by the public key for the mixer nodes. We can imagine this encryption in the form of Russian dolls with the message in the innermost layer. The message includes many layers and the mixer nodes by opening each layer, understand which node is the next destination. It is important to know that the mixer nodes open the layers by using their private keys. In the end, mixer nodes change the order of messages and send them to the next destination.

**2.3.3 Privacy in Mix network**

Mixer nodes can be vulnerable targets because they are attractive for malicious users. If malicious users penetrate in mixer nodes (for example, by injection of repackaged messages), they can damage the whole Mix network operation. In this section, we present several suggestions for improving privacy by introducing four dangerous threats to the privacy of Mixed Networks. We know that mixer nodes destroy links between senders and receivers. Therefore, malicious threats cannot identify users by analyzing the traffics of Mix networks. In case malicious users succeed in controlling some of the mixer nodes, they are not aware of previous message links, and they cannot identify the source and destination of the message. Mix networks can maintain anonymity in this way. The messages in the Mix networks contain unique information about the sender or receiver nodes.

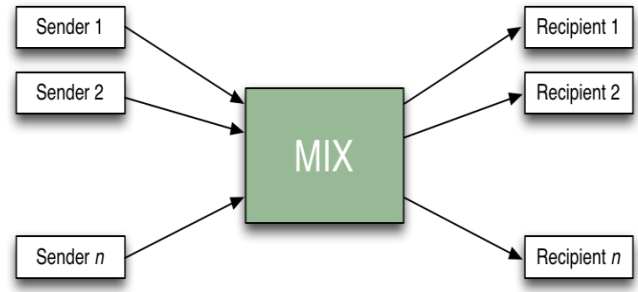

Figure 3. An example of Mix network. Adapted from [17]

By this information, the nodes identify each other better, which is an advantage of Mix networks to improve sending and receiving in their next forward messeges. Due to this, the networks know their nodes better in the next update. There is a negative aspect to this feature because if there is more information about nodes, the malicious users know more about the nodes' contacts, and they can analyze traffics of the Mix networks more easily and can break the anonymity and plan for more convenient attacks. To prevent this from happening, we need to limit the amount of the information we place in the Mix networks to only what is needed.

Each message in each mixer node is encrypted — this encryption made by a public key. Before each proxy is encrypted, this layer indicates which node is the next or which one is the final destination [16]. If a network does these actions poorly or does repeated actions in the proxy, malicious users can make predictions and compromise the privacy of the network users by analyzing the Mix networks' traffic and will find out about the connections between the network nodes. In such situations, malicious users monitor the arrival and departure of messages for a long time, or they test nodes that send messages from duplicate actions. In this way, they can predict the behaviour of nodes. The weakness in the transmitting process can also provide conditions for network penetration by malicious users. In this situation, malicious users can quickly create a template for prediction. They can use these weaknesses and, after penetration, break the privacy of the nodes and reach social graphs. To protect the networks, we must program proxies in a way that they do not allow duplicate actions or poor performance on encryption and decryption.

**3. THE PROPOSAL**

We plan to simulate a private and unlinkable exchange of messages by using a Public bulletin board and Mix network in Opportunistic networks. The goal is that the nodes can send and receive messages in an anonymous manner. In this simulation, the nodes can use paths which include some mixer nodes, and these mixer nodes can protect paths from being traced. This proposal can make sure that the security of the client runs in the end-user. Also, we simulate a Public bulletin board in which nodes can store their messages in a protected form. Senders can store their messages to the Public bulletin board, and the receivers can receive their messages by the Public bulletin board [18].

### 3.1 The Public bulletin board

As sketched in figure 4, we assume a Public bulletin board with n cells. The client can access to the Public bulletin board by using the Mix network. All cells in a Public bulletin board can contain a set of value/tag pairs. All cells are known to have an index that they have.

Users can operate two functions on a Public bulletin board. In the first function, the user can send an encrypted message with a preimage of the tag and an index to the Public bulletin board. A Public bulletin board can write encryption of a message in the value part of a cell. Also, a Public bulletin board can store preimage of the tag in the tag's part of a cell. The index of the cell in the Public bulletin board is chosen by the sender, and the Public bulletin board writes the message in the same index mentioned by the sender. It should be noted that the encrypted message must include a tag and an index for the next message from the sender.

In the second function, a receiver sends the preimage of a corresponding tag and an index — the preimage of the corresponding tag can open the Public bulletin board cells. We know that any node which is to be read in a Public bulletin board must know the corresponding tag of value. If a Public bulletin board can open the cell by the preimage of the corresponding tag, the value of the cell will return to the receiver; otherwise, it returns null. It should also be noted that only the intended recipients can delete the value of the cell by using the preimage of the corresponding tag.

We can adjust the size of a Public bulletin board in proportion to what a network needs. The size of the Public bulletin board can dynamically adjust with the network size so that we can reduce the amount of network activity and server activity.

Public bulletin boards can be attractive targets for malicious users. If malicious users can control Public bulletin boards, they can learn about all vital network information. It should be noted that the Public bulletin board does not know about the identity of senders and recipients. In a sense, the proposal will increase the level of security and privacy for Public bulletin boards.

### 3.2 Client protocols

As figure 4 shows, node A and B want to send a message to each other. They have met and reached an agreement before exchanging the first message. We suppose that nodes A and B shared a secret symmetric key in that meeting. After sharing the secret symmetric key, node B can use this key for opening the message from node A. It should be noted that both node A and B create a fresh key for each message. Each fresh key is made by a key derivation function. After receiving each message to node B, both node A and B update the symmetric key by key derivation function and delete the old copy of the key. We know that there has been an agreement between nodes A and B for the first index of a cell on the Public bulletin board and that node A gives the preimage of the first message.

As it was mentioned earlier, each cell has a number index in the Public bulletin board. Both node A and B can access a specific cell by the tag and index they share. Node B can read the value from a specific Public bulletin board cell because it recognizes the preimage of the corresponding tag and the first number index of that specific cell.

Figure 4 is an example of our simulation. There is no direct connection between nodes A and B when they send a message to each other. Instead, there are some usual and mixer nodes which can be used to create a path between node A and B.

Writing and reading from a Public bulletin board by nodes include two events during the process of simulation. During the first event, we assume that node A sends a message in a path that includes some mixer nodes and usual nodes. The message was sent along with a tag and index. The first mixer node receives the message and delivers it to the next mixer node, as long as the message and the rest of the content arrive in the Public bulletin board. Then the message will be stored in the same Public bulletin board cell according to the index number determined by node A.

For security reasons, it is better not to save messages in continuous cells in the Public bulletin board. Therefore, node A chooses a random index for the next message that refers to a new place in the Public bulletin board. This process can continue to work as long as the Public bulletin board is not filled.

In the second event, we suppose that the Public bulletin board receives a request message from node B. We know that nodes A and B have agreed about the first index and preimage of the corresponding tag. This message, along with the preimage of the corresponding tag and the first number index, will be read by the specific cell. The request message passes from a path with some mixer nodes, which makes it impossible to trace the path. After the request message arrives at the Public bulletin board, it checks the request message with the corresponding index and tag in the Public bulletin board cell. If the index of the tag was not empty, and the preimage of the corresponding tag can open the cell, the value of the cells returns to node B; otherwise, returns null. The Public bulletin board form the value of the cell as a new message and publish it in a path which includes some mixer nodes again.

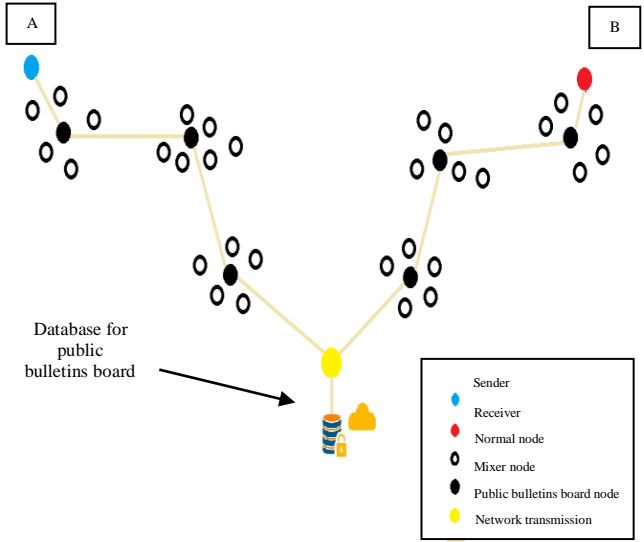

**Figure 4. An example of a Public bulletin board**

As we said earlier, a path with mixer nodes is protected from being traced. After the message passes along the path, it will arrive in node B along with the value.

After the arrival of the message (with a value) to node B, node B can decrypt the value of the message by the secret symmetric key shared before between node A and B. The information about the next index and tag mentioned in the encrypted value that node B can decrypt it now and know it. Node B decrypts the value and recognizes the index and tag of the next cell for a new message. Therefore, node B knows where the next message in Public bulletin board cells can be found. Also, node B can open the next cell by the tag it receives from the encrypted message of the value. Despite all the advantages, this proposal cannot say that this Public bulletin board is safe from any attack.

For example, we can consider a situation in which a malicious user sends many messages and fills all the cells in the Public bulletin board in his/her favour. A simple solution is that the Public bulletin board has registration for each node. It can prevent the Public bulletin board from attacks by a malicious node. By using this method, we can protect our public bulletin board form potential threats. Also, we suppose a value for each node, and the value decreases each time that a node stores a message in the Public bulletin board. Such actions could prevent these types of attacks from blocking the Public bulletin board.

## 4. Experimentation

### 4.1 Simulation environment

This proposal uses the ONE simulator for simulation— the ONE simulator is a java based simulator which is used for research in delay tolerant networks [19]. The ONE simulator can define different nodes with different behaviours and engage them in different scenarios.

The ONE simulator has some default functions which can be used to make, send, receive and route messages between nodes. In this simulator, we can create different scenarios by explaining different occurrences. We can also change the default functions of ONE simulator for specific routing protocols and design new protocols. We can change in the structure of messages and nodes too.

The ONE simulator can display outputs of simulation in two ways: graphically and textual. Our proposal has not used any graphics outputs; we have used textual outputs. We can customize outputs of the ONE simulator according to our proposal.

Also, the ONE simulator uses a function for making reports of simulation. The ONE simulator considers a Report file to gather all reports. These reports can provide enough information on how to work the simulation.

### 4.2 Scenarios

We define two scenarios in this proposal by using the ONE simulator. These two scenarios have some differences in speed movement of nodes and spatial dimensions of simulation. It should be noted that both scenarios have the same execution time.

In the first scenario, we assume a park in the dimension of one kilometer square and that forty-one nodes exist in this park. Each of these nodes is a human that can slowly walk with the speed of one to two kilometers per hour in the park and each node moves in a random direction. The number zero node is a Public bulletin board, and the size of it is one hundred cells. Other nodes can be normal nodes or mixer nodes. The range of the node's transmission is twenty-five meter, and the speed of the transmission is one hundred megabytes per second. Each node can use two prefixes for sending a message to the Public bulletin board: one prefix to write in the Public bulletin board and the second one to read from it. Each message can use three mixer nodes at maximum to reach the Public bulletin board.

In the second scenario, the Setting file is the same, except for the dimension of the park and the node's speed of movement. We assume a park as big as five hundred squares. Also, the node's speed of the movement is double time faster, and it is between two to three kilometres per hours.

### 4.3 The ONE modification

The ONE simulator contains various functions, each containing a set of different compiled java functions — each one designed for a specific purpose in the simulator. This proposal had the most changes in a series of java -functions in Core and Routing functions. Also, we created a new function in the core file which we called "BB". The BB function used to create a Public bulletin board. We also needed to make changes to the Routing functions so that we could define a protocol between senders and receivers with a public bulletins board. Also, this simulation considers Mix network function, which can change each node to become a mixer node. This section only discusses changes in the functions and the creation of BB java function.

We should know that after all changes on the ONE simulator (which will be explained), we can compile all the functions to record the changes. After running the simulator on our desired setting on Setting files according to our proposal, the ONE simulator can execute the simulation. Also, in the terminal, we can see outputs of each step of sending, receiving, and storing messages from the Public bulletin board. After the occurrence of each step, the status of the Public bulletin board will be shown in the terminal.

### 4.3.1 Public bulletin board function

In the first step, we write a java function for creating a Public bulletin board structure. We create this function in core functions.

The structure of the Public bulletin board contains two arrays, one array for storing the preimage of tags and another for storing values. Also, each cell of arrays has a number index by which we can access the cell.

The Public bulletin board function has two main methods. One of the methods is the WRITE method, which is used to write the value of messages and the preimage of tags from senders, and another one is the READ method to return values from a Public bulletin board to receivers.

The WRITE method is the first method that we considered in the java Public bulletin board function (BB). At first, when a message arrives in the Public bulletin board, BB function checks the message's prefix. If the message's prefix is equal to the WRITE prefix, the WRITE method becomes active. Then, the WRITE method starts to store the message in the Public bulletin board arrays. As we know, a message includes a value, the preimage of tag and an id. The message's id refers to the numbered cell of arrays. The WRITE method receives the message's value and stores it in the value array. The cell's number of the array is equal to the message's id. The WRITE method does the same with the preimage of the message's tag and stores it in tags array.

The READ method is the second method we considered in the java Public bulletin board function (BB). After checking the prefix message by the Public bulletin board function, if the message's prefix is equal to the READ prefix, the READ method becomes active. It should be noted that a message to get value from the Public bulletin board needs an id and a corresponding tag. The READ method checks the cell's number of the tags array according to the message's id. The READ method checks the value array with the same number of id to see if the cell's tag is equal to the corresponding message's tag. Then the READ method returns the value in the form of a new message. -The READ method sends the new message in a path that including mixer nodes to reach the leading destination by using routing function.

### 4.3.2 The DTNHost function

In the ONE simulator, all nodes are created by DTNHOST function. The structure of a node created by the DTNHOST function consists of the name, location, type of movement, range of transmission, speed of transition, etc. Even we can by call BB function in a host structure and use a Public bulletin board structure in it. Also, we add three values, including a secret key, an agreed-id and an agreed-tag in DTNHOST function. It should be noted that we wrote these three functions to generate the secret key, the agreed-id and the agreed-tag that it takes amount randomly for each one. Also, we write methods to generate their value.

### 4.3.3 The SIM Scenario function

Scenario function can create a scenario that is supposed to happen according to Setting files. We have modified the Scenario function. In a part of this function that takes the number of the host in simulation from the Setting function, we can have constructed the first node as the Public bulletin board to create nodes in the simulation when it uses DTNHOST function.

### 4.3.4 The Message function

The Message function is used for creating message structures in the ONE simulator. The message structure has values including sender hosts, receiver hosts, the id of the message, the size of the message, etc. We add some values for carrying the agreed-id, the agreed-tag and the encrypted value. Also, we have created a method that works automatically when a message is generated. This method simulates the agreement between sender hosts and receiver hosts. When a message is created automatically, the value of the secret key, the agreed-tag and the agreed-id between a sender host and a receiver host becomes equal. Also, the id and tag in the message structure become equal with the agreed-id and the agreed-tag values from the sender.

### 4.3.5 The Mix network function

We create a java function in the routing functions named Mix network function which generates the mixer nodes. As mentioned in the previous section, using mixer nodes in the simulation can create an untraceable message which has been one of the goals of the present proposal. At each mixer node, messages become mixed without any change to their context.

### 4.3.6 The Message routing function

The message routing function introduces a protocol to route a message from senders to receivers. To do so, we should have some changes on message routing function to adapt simulation with the Public bulletin board. Firstly, in Message routing function, we check the message destination to make sure whether or not it exists in the Public bulletin board. If the destination exists in the Public bulletin board, we activate the Public bulletin board function to check the message. Also, as we know, a Public bulletin board uses Message routing function to send new messages it generates to return the values. After generating a new message by the Public bulletin board, we change the address of the host who wants to receive a message of the Public bulletin board with zero node address. Then we add the new message in the list of the message for routing in simulation. New messages will use mixer nodes to reach the leading destination. Also, after the routing of all new messages, we can access the information about all new messages in reports files.

### 4.4 Results

#### 4.4.1 Metrics to be studied

We review the result of the simulation by considering three metrics that include Total packet delivery ratio, Total packet latency average and Total packet latency median.

The Total packet delivery ratio is the ratio of the total packets successfully sent to the Public bulletin board. We calculate the Total packet delivery ratio by the sum of the ratio delivery of all packets that arrive at the Public bulletin board (all packets that use one, two, or three mixers or do not use any mixers). Then, we divided the total delivery ratio by four. We show the Total packet delivery ratio in Figure 5 and Figure 8 for each scenario by two columns in blue and red. The blue column refers to nodes which store messages in the Public bulletin board, and the red column refers to the nodes which receive messages from the Public bulletin board.

The total packet latency average refers to the total delay of packets that arrive at the Public bulletin board. We calculate the total packet latency average by the sum of the latency of all packets that arrive at the Public bulletin board, packets that use one, two, or three mixers or do not use any mixers. Also, in Figure 6 and Figure 9, we consider two columns with red and blue colours.

The total packet latency median refers to the total delay of packets to arrive at the Public bulletin board for both methods. We calculate Total packet latency median by the sum of the median of all packets that arrive at Public bulletin board, all packets that use one mixer or two mixers or three mixers or do not use any mixer. Also, in Figure 7 and Figure 10, we consider two columns with different colours as previous figures.

#### 4.4.2 Result

For the first scenario, we provided three charts. Figure 5 refers to the Total packet delivery ratio. In both methods (READ with the red column, WRITE with the blue column), the results of the Total packet delivery ratio have an acceptable ratio. Therefore, we can understand that if a node sends a message for storing in the Public bulletin board, this message reaches the destination with a high rate. Also, in Figure 6 and Figure 7, we can see the Total packet latency average and the Total packet latency median for the total number of messages for each method. Unfortunately, if we look at these figures, we notice that the simulation has not worked very well in terms of network latency and median. A message should stay quite a few minutes in intermediate nodes to reach the destination, which is a long time for the delivery of a packet, as we know. If our Setting file includes the slow movement of the nodes and the size of the simulated environment, we can justify these results. In general, these results are not acceptable.

As we said, we had some changes in the Setting file for the second scenario. In the following next three charts, we obtained results by doubling the speed of node movement and half of the simulation park dimension but with the same execution time. Figure 8 shows the Total packet delivery ratio for the second scenario. At first, the rate for the second scenario is very acceptable as the first scenario. It can be a reasonable rate for simulation to deliver each message. Figure 9 and Figure 10 show the Total packet latency average and the Total packet latency median. An important point in Figure 9 and Figure 10 is the Mix network condition (because of the messages stored in the mixer nodes many times), which makes the latency average, and latency median acceptable. In general, these results are good results for the second scenario.

#### 4.4.3 Conclusions on the results

By considering all the charts, we can compare two scenarios. Based on the Total packet delivery ratio of both scenarios, we can see that our simulation has a high performance in terms of delivery rates. In general, we can conclude that the simulation has a good Total packet delivery ratio.

As we have seen in the first scenario, the latency time rate was very high; it is not an acceptable rate for our simulation. Moreover, with the changes we made in the second scenario, we achieved acceptable results in terms of network latency and median. We know that we had a small modification in the second scenario, but we can see a massive difference between the charts. Thus, the simulation performance is related directly to the node's speed of movement and the size of the simulation environment.

By doubling the speed of the nodes, we reduce the delay of the message several times. We can say that the simulation in the second scenario has a good result by considering the Total packet latency average and the Total packet latency median. If we look at our simulation performance by considering all the charts, it can be concluded that our simulation can work well with a central Public bulletin board for environments where nodes have a faster speed of movement. Given the overall results, it can be concluded that the simulation has a reasonably good performance.

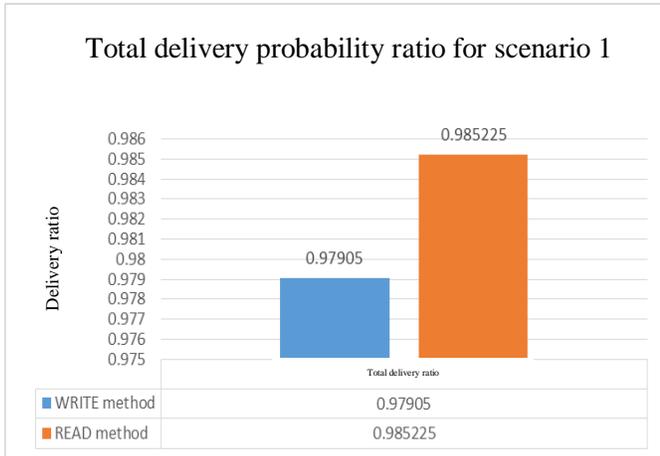

**Figure 5.** Total delivery probability ratio for scenario 1

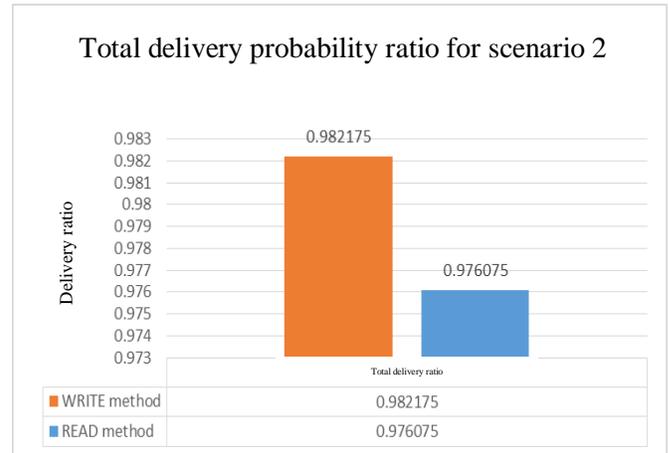

**Figure 8.** Total delivery probability ratio for scenario 2

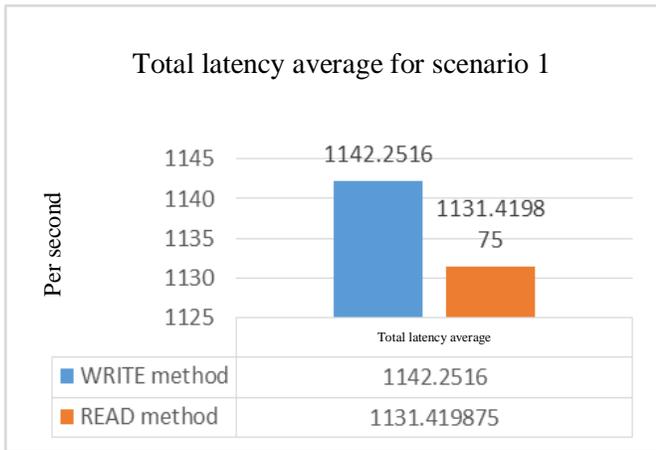

**Figure 6.** Total latency average for scenario 1

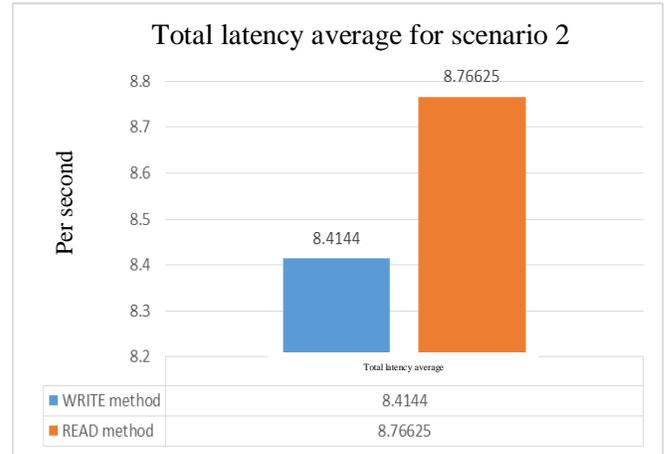

**Figure 9.** Total latency average for scenario 2

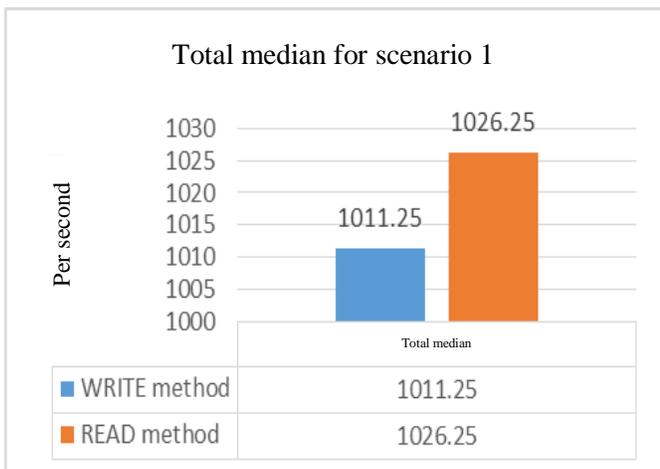

**Figure 7.** Total latency median for scenario 1

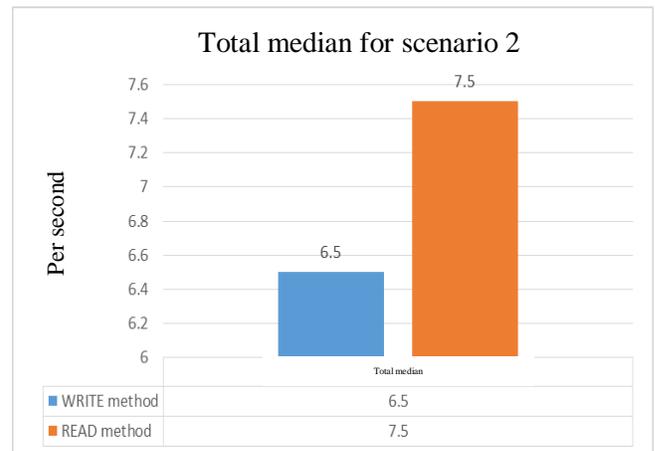

**Figure 10.** Total latency median for scenario 2

**5. CONCLUSIONS**

As we discussed in the document, we intended to simulate a private and unlinkable exchange of messages by using a Public bulletin board and Mix networks in opportunistic networks. We used the Mix network, where the nodes could send their messages through intermediate nodes anonymously. By considering the materials from the related work section, we provided a protocol in which nodes could send their packets to the public bulletin in a secure environment. Also, we considered two different scenarios, and we analyzed and compared them.

If we want to draw a general conclusion about the results of the simulation, we can say that our simulation has a satisfactory performance. Nodes can operate and use the benefits of the public bulletins board in our simulation. Also, the nodes can send their packets in anonymity by using Mix network. At all stages, we succeeded in preserving the anonymity of the nodes. In general, we can say that this proposal was able to achieve overall goals at an acceptable level.